**True superconductivity at near ambient temperature has not been confirmed by Dasenbrock-Gammon, et. al. Nature, volume 615, pages 244–250 (2023).**

**Comment on Dasenbrock-Gammon, et. al. Nature, volume 615, pages 244–250 (2023). "Evidence of near-ambient superconductivity in a N-doped lutetium hydride"**

Srinivas Sridhar
Department of Physics, Northeastern University, Boston, MA 02115.
s.sridhar@northeastern.edu

## Abstract

The paper by Dasenbrock-Gammon, et. al. Nature, volume 615, pages 244–250 (2023), does not present evidence for true superconductivity at near ambient temperature in a N-doped lutetium hydride system.

In a recent paper in Nature, Dasenbrock-Gammon, et. al.[1] have shown some intriguing results of a resistance transition at temperatures as high as 294K at high pressures in a lutetium–nitrogen–hydrogen system. The authors claim zero resistance and increased diamagnetism measured by vibrating magnetometry below the transition temperature.

While these results are very intriguing and potentially exciting, they do not confirm true superconductivity below the transition temperature.

The hallmarks of true superconductivity[2] are well known and are summarized below.

#1: **Zero resistance demonstrated by persistent currents**. The authors use a 4-terminal resistance measurement which is not a true measure of intrinsic resistivity. Their results only demonstrate that the measured voltage drops below the sensitivity of their voltmeter and is not a true test of zero resistance. The ultimate test of true zero resistance would be the demonstration of persistent currents in a ring of the material with almost negligible decay over observable laboratory time scales.

#2. **The Meissner effec**t[3], evidence of bulk diamagnetism at zero frequency. The authors present data for increased diamagnetism below the transition temperature as measured by a vibrating magnetometer setup. The data are incorrectly presented as "dc magnetization". Since a vibrating magnetometer measurement is carried out at finite frequency, it is an inductive measurement and not a true dc magnetization measurement. Thus a transition to lower resistance will also lead to an apparent diamagnetism in an inductive measurement due to increased induced shielding currents. The resistive and magnetic transitions are truly decoupled only at zero frequency. Only a true dc susceptibility measurement (such as with a SQUID magnetometer) could provide evidence for a Meissner effect. The magnetic susceptibility data presented in the paper do not conclusively demonstrate a true Meissner effect arising from the onset of superconductivity.

The Josephson effect[4], which provides evidence for a macroscopic condensate, can be sometimes used as evidence of interfacial or granular superconductivity[5]. No data on the Josephson effect are presented in the paper.

Instead, the data presented are consistent with increased metallicity occurring below an apparently sharp transition temperature, and do not distinguish between a magneto-resistive or superconducting transition. While these data may represent a potentially novel phenomenon in metallic systems at high pressures, they do not meet established criteria for demonstration of a superconducting material below the transition temperature.

---